\documentclass[11pt]{article}  
\usepackage{graphicx}  %
\usepackage[numbers,sort]{natbib}
\usepackage{float}
\usepackage{amsmath}  
\usepackage{amssymb}  
\usepackage{authblk}

\title{Causal Emergence 2.0: Quantifying emergent complexity}
\author[1]{Erik Hoel\thanks{\texttt{erik.hoel@tufts.edu}}}
\affil[1]{Allen Discovery Center, Tufts University, Medford, MA, USA}

\begin{document}

\maketitle

\begin{abstract}

Complex systems can be described at myriad different scales, and their causal workings often have multiscale structure (e.g., a computer can be described at the microscale of its hardware circuitry, the mesoscale of its machine code, and the macroscale of its operating system). While scientists study and model systems across the full hierarchy of their scales, from microphysics to macroeconomics, there is debate about what the macroscales of systems can possibly add beyond mere compression. To resolve this longstanding issue, here a new theory of emergence is introduced, wherein the different scales of a system are treated like slices of a higher-dimensional object. The theory can distinguish which of these scales possess unique causal contributions, and which are not causally relevant. Constructed from an axiomatic notion of causation, the theory's application is demonstrated in coarse-grains of Markov chains. It identifies all cases of macroscale causation: instances where reduction to a microscale is possible, yet lossy about causation. Furthermore, the theory posits a causal apportioning schema that calculates the causal contribution of each scale, showing what each uniquely adds. Finally, it reveals a novel measure of emergent complexity: how widely distributed a system's causal workings are across its hierarchy of scales.

\end{abstract}

\section{Introduction}

Complex systems operate across scales, and therefore evince a huge number of possible descriptions \cite{west2018scale}. This has led to claims that there is “no privileged level of causation” in complex systems, such as in biology \cite{noble2012theory}. However, the set of all possible scales---in the form of dimensionality reductions---is extremely large, even for small systems, and the majority are poor descriptions of a system's causal workings (e.g., randomly coarse-graining the logic gates of a computer).

This embarrassment of multiplicity necessitates a formal mathematical theory of emergence. Such a theory should explain and quantify how macroscales (higher-level descriptions of a system, based on some dimensionality reduction) contribute to a system's causal workings, and which macroscales are causally relevant. A theory of emergence may even explain the spatiotemporal hierarchy of the sciences themselves, beyond their function of just useful compressions \cite{hoel2024world}.

It might be protested that there is no room for emergence in science, as presumably the future of any given system can be predicted with full knowledge of its microscale, and presumably any given system can be reduced to its microscale. However, prediction is not the same thing as causation \cite{grasso2021causal}. A toy example is that of a thermostat and room system \cite{hoel2018agent, flack2017coarse}. While in theory the microscale of all the individual particles in the room could be used to predict the thermostat's reading, in terms of causal understanding it represents a poor answer to the question of “What caused the thermostat to read $20^\circ \text{C}$?” In fact, the exact microstate of all the particles is not causally necessary for the reading, since many other configurations could lead to it. Meanwhile, the macrostate (the temperature of the room) has a direct causal relationship to the thermostat's reading, in that it is necessary for any given value.

In another example, an incoming signal to a neuron's dendrites could be used to predict that a downstream action potential will occur. Yet, in a causal analysis, the incoming signal would be insufficient to trigger (as an effect) some exact exchange of ions, as this would evolve unpredictably due to noise, such as from Brownian motion \cite{braun2021stochasticity} or quantum effects \cite{jedlicka2017revisiting}. Meanwhile, the incoming signal could still be deterministically sufficient to trigger (as an effect) the downstream neuronal macrostate of “firing," irrespective of its underlying microscale details.

Following these intuitions, in 2013 my co-authors and I introduced the theory of causal emergence \cite{hoel2013quantifying}. The theory made use of discrete causal models (in the class of logic gate networks, DAGs, and Markov chains), and a measure of causation, the effective information (EI) \cite{hoel2017map}. The theory provided a toolkit to search across all possible dimension reductions of such systems to find the one that maximized the EI (wherein the EI is assessed by perturbing the system into all possible states via the \textit{do}(\textit{x}) operator \cite{pearl2009causality}, and then calculating the mutual information between that probability distribution of interventions and the probability distribution of their effects). Identifying the macroscale with the maximal increase in EI quantified the degree of causal emergence in the system.

Causal emergence revealed why macroscales of a system can have stronger causal relationships despite being reducible to their underlying macroscales: since macroscales are multiply-realizable, they can minimize the uncertainty in causal relationships, which a measure of causation like the EI is sensitive to. This is mathematically similar to how coding over an information channel can minimize the noise of communication \cite{hoel2017map, hoel2024world}.
    
The theory of causal emergence has since spawned a large amount of research, such as measuring causal emergence in data spanning from cellular automata \cite{varley2020causal} to fMRI data \cite{yang2025finding} to gene regulatory networks \cite{pigozziassociative, hoel2020emergence}, as well as developing heuristics \cite{griebenow2019finding}, like detecting causal emergence with trained artificial neural networks \cite{zhang2022neural}. It's been related to phenomena like scale-freeness and robustness in network theory \cite{klein2020emergence, klein2021evolution}, been adapted within Integrated Information Theory \cite{hoel2016can, marshall2024micro}, and there have been proposals for alternatives of what measure should be used for quantifying causal emergence, such using the dynamical reversibility of a system to approximate the EI \cite{zhang2025dynamical}. For a full review, see Yuan et al. 2024 \cite{yuan2024emergence}.

However, the initial version of causal emergence (hereafter, CE 1.0) has remained incomplete, due to two outstanding issues. The first is the reliance on the EI (and its approximations) to detect causal emergence. While the EI is a relatively well-constructed measure of causation \cite{balduzzi2011information}, it makes background assumptions in its calculations (such as requiring a uniform distribution of interventions, which some have criticized) \cite{eberhardt2022causal, aaronson_higher-level_2017, comolatti2022causal}. Additionally, as demonstrated here in Section 6.2, the use of EI actually underestimates causal emergence.

The second issue is that CE 1.0 only identified a single causally-relevant scale (the maximum of EI), ignoring all multiscale structure. Yet many systems seem to operate across scales; a prominent example being the brain's different functional scales, ranging from the neuronal up to cortical minicolumns up to entire brain regions \cite{yuste2015neuron}. Another example is how a computer can be described at the microscale of its hardware circuitry, the mesoscale of its machine code software, or at the macroscale of its operating system and applications \cite{rosas2024software}; indeed, even what computations are occurring may change depending on the scale of description \cite{gomez2022slicing, bongard2023there}.

To develop a universal and well-grounded theory of causal emergence that resolves these issues, here I introduce a novel formalization: Causal Emergence 2.0. The fundamental intuition of CE 2.0 is that a system is not bound to one particular scale of description; rather, it is best described by the set of scales that contribute to the system's causal workings. Any one scale (even the microscale) is much like taking a 2D slice of a 3D object, and therefore cannot fully capture the causation of the system. CE 2.0 posits a \textit{causal apportioning schema} across scales that detects their causal contributions (if any) to the multiscale whole. This is accomplished via defining a path that traverses a system's scales from “top to bottom,” and the theory apportions out the causation of a system's workings along this path.

CE 2.0 is grounded in an axiomatic notion of causation, rather than a stand-alone measure (like the EI in CE 1.0), which allows the theory to capture all cases of macroscale causation, and also unfold and quantify the multiscale causal structure of systems in ways previously impossible. This new taxonomy of how systems operate across scales leads to a novel measure, the \textit{emergent complexity}: how widely distributed a system's causal workings are across its scales, wherein systems that possess many contributing scales are more complex.

In what follows, CE 2.0 is outlined, first by defining a notion of causation that is axiomatic and robust across background assumptions, then by using that to calculate the degree of macroscale causation in coarse-grains of model Markov chains and therefore to quantify their degree of causal emergence, then by detailing how causal contributions are assigned via a path that traverses the set of scales, and furthermore exploring how this naturally arrives at the notion of emergent complexity. Finally, CE 2.0 is directly compared to other related theories of emergence, demonstrating its advantages and outlining its conceptual implications.
    
\section{The axioms of causation}

\subsection{Sufficiency and necessity}

Scientists distill and extract causal knowledge about the systems they study \cite{beebee2009oxford}. Breakthroughs in the scientific understanding of causation include things like R.A. Fisher's formalization of randomized controlled trials \cite{fisher1999advances}, as well as Judea Pearl's more recent introduction of the \textit{do}(\textit{x}) operator \cite{pearl2009causality}.

A number of researchers have introduced specific probabilistic measures of causality to capture the degree of causation between a cause and an effect. The aim of such measures can be described in various ways, e.g., as capturing the power of a particular cause, the strength of a particular causal relationship, the amount by which one variable causally controls another, etc. Applying such measures of causation involves specifying a causal model, then using counterfactuals \cite{lewis1973causation} or interventions \cite{pearl2009causality} to separate causal knowledge from mere observation.

Recent work analyzing over a dozen proposed measures of probabilistic causation by different authors \cite{comolatti2022causal} showed that in the scientific literature there is \textit{causal consilience}: measures of causation, independently introduced across fields from psychology to statistics to philosophy \cite{fitelson2011probabilistic}, all set in relation two basic terms. These terms were dubbed “causal primitives” \cite{comolatti2022causal}---more commonly, they are known as the \textit{sufficiency} and the \textit{necessity}. Consilience held true for measures of causation ranging from those proposed by philosopher David Lewis \cite{lewis1973causation}, to to mathematician Judea Pearl \cite{pearl2009causality}, to my and my co-author's recent definition of actual causation \cite{albantakis2019caused}. Rediscovered many times independently, the primitives form an axiomatic foundation for any measure of causation, and ensure that measures of causation have significant overlap in their mathematical behavior. Ultimately, each term represents an inverse of uncertainty: sufficiency is the certainty about an effect, given a cause, whereas necessity is the certainty about a cause, given an effect.

As will be shown, the causal primitives of sufficiency and necessity have a further information-theoretic generalization, the determinism and degeneracy (respectively). CE 2.0 is based explicitly on these primitives and their further generalization.

First, to define the causal primitives formally, some terminology is required. For a discrete system like a Markov chain, DAG, or set of logic gates, assessing the causal primitives (hereafter CP) involves specifying some abstract space associated with the system, $\Omega$, which defines the set of occurrence (e.g., states, or events, variables, etc.) entered into the causal analysis. Then, for any occurrence (like a state transition), we can define the potential causes, $C\in\Omega$, and the potential effects, $E\in\Omega$. 

As the theory will be specified in simulated Markov chains, here $\Omega$ is simply a system's statespace. Occurrences are then a state transition, in that for a Markov chain there is always some individual preceding state of the system, $c$, and its effect $e$, the next state. These transitions each have some probability, \textit{P}.

Given an occurrence (here, a state transition), the sufficiency of a cause is then

$$ \text{\textit{suff}}(e,c) = P(e \mid  c), $$

which increases as $c$ is more probabilistically likely to bring about $e$, and reaches 1 when $c$ is fully sufficient for $e$.

The necessity of a cause is

$$ \text{\textit{nec}}(e,c) = 1 - P(e \mid C, \neg c),$$

which specifies the probability of $e$ occurring without $c$. That is, given some set of causes $C$ within the system, and given that, within that set, $c$ itself did not occur, what is the inverse of the probability of $e$? This probability is low when there are many common causes, and is $1$ only when $c$ is absolutely necessary for $e$, in that no other members of $C$ could produce it (in a Markov chain, this would mean no other states lead to the state $e$ at the next timestep).

Since the goal is to compare causation between entire scales, the terms “sufficiency” and “necessity” (and their joint description as CP) will henceforth imply their system-wide average across all $t$ to $t_{+1}$ transitions. This average is weighted by the probability of each transition $P(e \mid c)$, given $P(C)$. 

In a causal analysis, $P(C)$ is purposefully not the observed distribution \cite{pearl2018book}. The choice of $P(C)$ can be conceptualized as identifying the set of viable counterfactuals or, alternatively and identically, specifying some set of possible interventions in the causal model of a system \cite{pearl2018book, hoel2017map}. Since the following values will be calculated in Markov chains defined by some transition probability matrix (TPM) and its transitions (from $t$ to $t_{+1}$), here $P(C)$ is a uniform distribution across the entire set of states, given some scale of a system.

Conceptually, this just means that states of a scale are treated equally when it comes to being viable interventions or counterfactuals to assess the causal relationships of other states. E.g., given a COPY gate with a self-loop of $p=1$, this would imply we consider both 0 and 1 equally in $P(C)$, and therefore could correctly say that the state COPY $=1$ (at $t$) is sufficient and necessary for the state COPY $=1$ at $t_{+1}$ (note that with an observed distribution such a judgment would be impossible).

\subsection{Determinism and degeneracy}

The sufficiency and necessity each have an information-theoretic generalization: the \textit{determinism} and the \textit{degeneracy} of a system \cite{hoel2013quantifying}. 

Specifically, the determinism is the inverse of noise (or randomness) in the probability distributions of state transitions, and therefore is an information-theoretic generalization of the sufficiency. For an individual cause $c$ it can be defined as a coefficient, based on the entropy of $c$'s transition probability distribution over the set of effects $E$ in the system (normalized to range from $[0, 1]$):

$$ \textit{determinism} = 1 - \frac{H(E\mid c)}{\log_2 n},$$

wherein the central entropy term is

$$ H(E\mid c) = \sum_{e \in E} \text{\textit{suff}}(e,c) \log_2\frac{1}{\text{\textit{suff}}(e,c)}.$$

In what follows, the term “determinism” is reserved for the system-wide determinism coefficient, which is the average of the above determinism of an individual cause, given $P(C)$. Determinism is maximal only if the TPM consists of "one hot" rows, and is zero only if all rows are uniform distributions (i.e., transitions are random).

“Degeneracy” is also a system-wide degeneracy coefficient, defined as the inverse of the entropy over the probability distribution of the full set of effects, $P(E)$, given $P(C)$:

$$ \textit{degeneracy} = 1 - \frac{H(E \mid C)}{\log_2 (n)}. $$

Degeneracy is high if causes have many similar effects. Degeneracy is zero when all causes have a unique effect. In Markov chains, a fully deterministic and non-degenerate system would be one in which every state transitions with $p=1$ to some unique next state, with zero overlap (a permutation matrix). Degeneracy acts as an inclusive inverse of the necessity, in that

$$H(E\mid C) = \sum_{e \in E}P(e \mid C)\log_2\frac{1}{P(e \mid C)}, $$

wherein $P(e \mid C)$ is the inclusive form of the central term of the necessity calculation $P(e \mid C, \neg c)$. Instead of removing the cause ($\neg c$), it calculates, for any given $e$, how necessary the possible causes that lead to it are overall, out of the full set of $P(C)$.

In order to avoid linguistic confusions around inverses (since a low degeneracy indicates a stronger causal relationship), here during calculations the degeneracy is often reversed into the \textit{specificity}, for which increasing values indicate stronger causal relationships:

$$ \text{\textit{specificity}} = 1-\textit{degeneracy}. $$

To summarize: as can be seen by their construction from the same basic probability terms, determinism is essentially the normalized entropy of the sufficiencies and degeneracy is essentially the normalized entropy of the necessities (as its inverse, and calculated inclusively). 

The causal primitives and their generalizations are sensitive to the uncertainty in a causal model's relationships regarding causes and effects, and behave similarly when uncertainty is decreased or increased, making them suitable foundations for detecting macroscale causation. In S1 in the SI, their sensitivity and similarity are demonstrated via simulations of probability redistribution within a system.

\section{Quantifying macroscale causation in CE 2.0}

CE 2.0 defines macroscale causation as when the causal primitives (CP) are jointly greater at a macroscale of a system, indicating that the macroscale has reduced the uncertainty about causes and effects. Note that not all dimension reductions result in macroscales reflecting gains in CP---many lead to zero gains or even decreases (cases of causal reduction). To identify positive gains, CE 2.0 makes use of an ordered micro$\rightarrow$macro path that traverses the hierarchy of scales of a system, revealing its multiscale structure. The degree of causal emergence in a system is the total gain in CP along a micro$\rightarrow$macro path, representing the sum of macroscale causation across all scales. This causation is then apportioned along the path to track which scales have positive causal contributions (the degree to which they increase CP).

\subsection{Traversing the hierarchy of scales with a micro$\rightarrow$macro path}

Macroscales are defined in CE 2.0 as the result of some dimension reduction of a system. For a given Markov chain, $S$, and its associated TPM, which represents the microscale, some new system $S_M$ is defined with its own associated TPM, with macrostates replacing a set of microstates, wherein the transitions to and from a given macrostate are a summary statistic of the underlying microstates.

For simplicity, here I only consider dimension reductions that are coarse-grains of microstates. Coarse-grains result in macroscales like $(0, 1, 2), (3)$, which would indicate that for some 4-state system the microstates $(0, 1, 2)$ have been coarse-grained into a macrostate, while $(3)$ remains as it was at the microscale. This method follows previous research on causal emergence (for full details on how to derive a macroscale's TPM based on a coarse-graining of the microscale's TPM, see \cite{hoel2013quantifying, klein2020emergence}).

All coarse-grains were checked to ensure their validity as accurate descriptions of their underlying microscale; specifically, much as in \cite{klein2020emergence}, the dynamical consistency of each macroscale was checked, and inconsistent macroscales were discarded (see S2 in the SI). Note that, CE 2.0 can also be applied over other types of dimension reductions similar to coarse-graining, like black-boxing \cite{marshall2018black, hoel2017map} or higher-order macrostates \cite{klein2020emergence}. 

A micro$\rightarrow$macro path is the set of valid (i.e., dynamically consistent) scales that lead from the microscale up to a final macroscale, which acts as the endpoint of the path. Conceptually, a path is simply specifying what coarse-grains of the system are “on the way” to other coarse-grains across the hierarchy of scales that spans the system. Each step in the path is a single “slice” of the set of all possible scales, which the micro$\rightarrow$macro path traverses from bottom (full dimensionality) to top (final dimension reduction).

In a hypothetical 4-state system, the coarse-grain of $(0, 1), (2), (3)$ would be on a path to the lower-dimensionality coarse-grain of $(0,1,2), (3)$. Along such a path, the microstates $(0)$ and $(1)$ would first be coarse-grained together into a single macrostate, $(0,1)$, and then further coarse-grained into $(0,1,2)$. A full micro$\rightarrow$macro path for such a hypothetical 4-state system, starting at a microscale, might be $(0), (1), (2), (3)$ $\rightarrow$ $(0, 1), (2), (3)$ $\rightarrow$ $(0,1,2), (3)$ $\rightarrow$ $(0,1,2,3)$, ending with all states coarse-grained into one macrostate. Formally, this is just:

\[
\pi^{(1)} \;\longrightarrow\; \pi^{(2)} \;\longrightarrow\; \cdots \;\longrightarrow\; \pi^{(k)},
\]

wherein each $\pi^{(i)}$ is some valid partition (representing a coarse-grain) of the original $n$ microstates, and $\pi^{(i+1)}$ is a coarse-grain in turn of $\pi^{(i)}$, concluding at the endpoint, partition $\pi^{(k)}$.

As an example, a chosen micro$\rightarrow$macro path is plotted for the 8-state Markov chain visualized in Figure 1. Progress along the path is shown in Figure 1 via color contagion. Beginning at $(0),(1),(2),(3),(4),(5),(6),(7)$, the microscale, microstates coarse-grained together along the path are changed to be the same color at each coarse-graining step until the endpoint of $(0),(1,2,3,4,5,6,7)$, wherein all microstates have been coarse-grained together, except $(0)$.

\begin{figure}[H]
    \centering
    \includegraphics[width=1\textwidth]{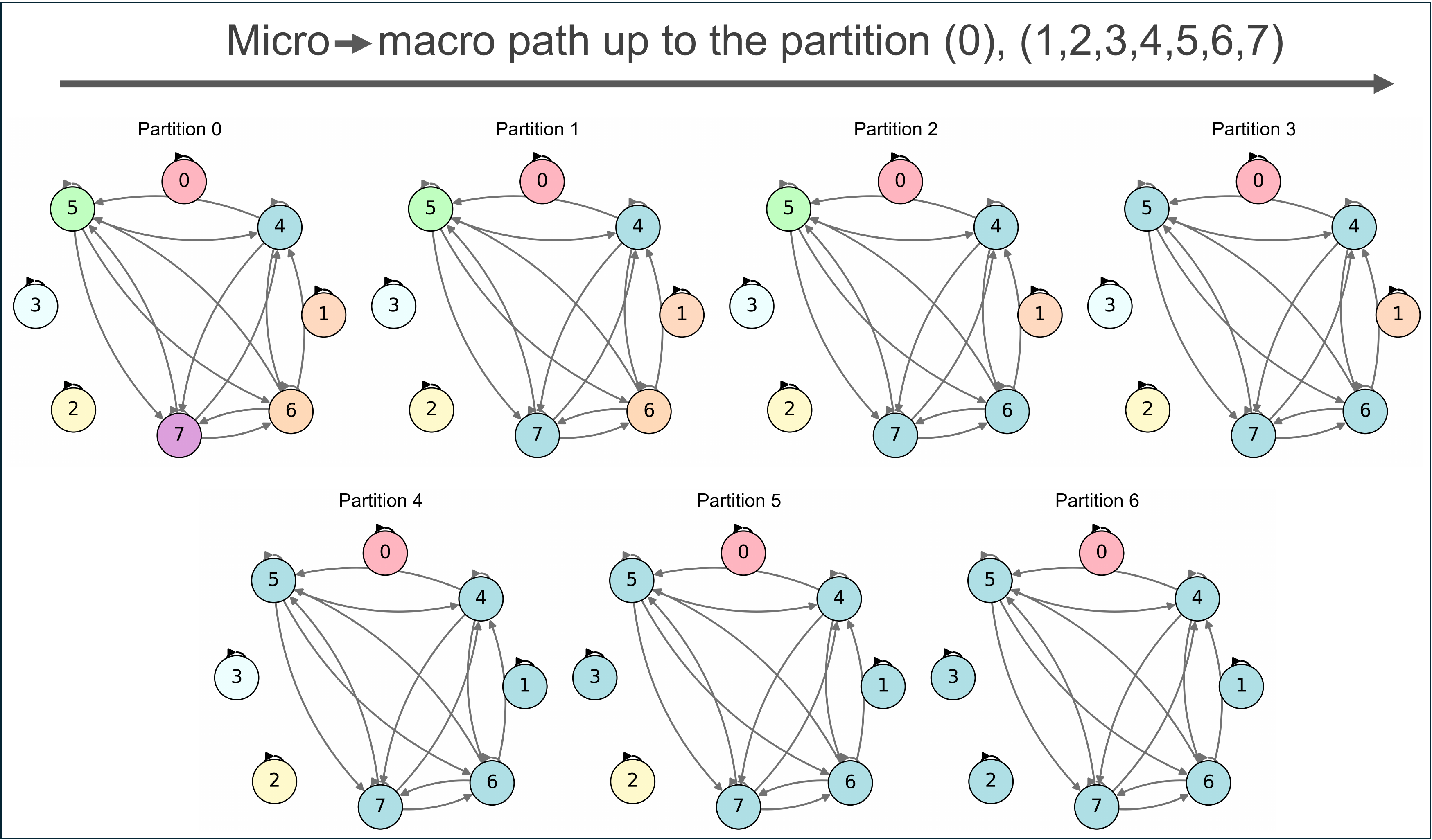} 
    \caption{\textbf{A micro$\rightarrow$macro path visualized}. An example 8-state Markov chain, with the probabilities of transitions represented in grayscale (for its TPM, see Fig.2A). Starting at the full partition of $(0), (1), (2), (3), (4), (5), (6), (7)$ (the microscale) states are coarse-grained together, with each further partition being a step in the path (and thus a scale in the system), ending at $(0), (1, 2, 3, 4, 5, 6, 7)$. Changes to being the same color indicate when states are coarse-grained together along the chosen path (color contagion).}
    \label{fig:3}

\end{figure}

\subsection{Causal apportioning along a path}

The distributions of gains in CP due to dimensionality reduction, calculated at each scale along a path, can be found via a \textit{causal apportioning schema}. Specifically, for a system and a chosen micro$\rightarrow$macro path, CE 2.0 calculates \(\Delta \mathrm{CP}\) at each step in the path, compared to the previous scale.

Specifically, for a given scale (each step in the path) the CP value is the sufficiency and the necessity of the macroscale (or microscale, at the start of the path) added together---then 1 was subtracted from this value, to provide a bound of $[0,1]$. Just the same, the CP value of the determinism and specificity was also calculated at each step by adding them together and subtracting 1 (making this value equivalent to the \textit{effectiveness} from CE 1.0 \cite{hoel2013quantifying}).

Calculating the \(\Delta \mathrm{CP}\) along a path is demonstrated in an 8-state Markov chain, the starting microscale TPM of which is shown in Fig. 2A. It has an intuitive macrostate in the form of an equivalency class across microstates $(4, 5, 6, 7)$, which all share identical probability distributions of transitions. Its connectivity (state transitions) is the same as visualized in Figure 1, and the same chosen micro$\rightarrow$macro path is used.

\begin{figure}[H]
    \centering
    \includegraphics[width=1\textwidth]{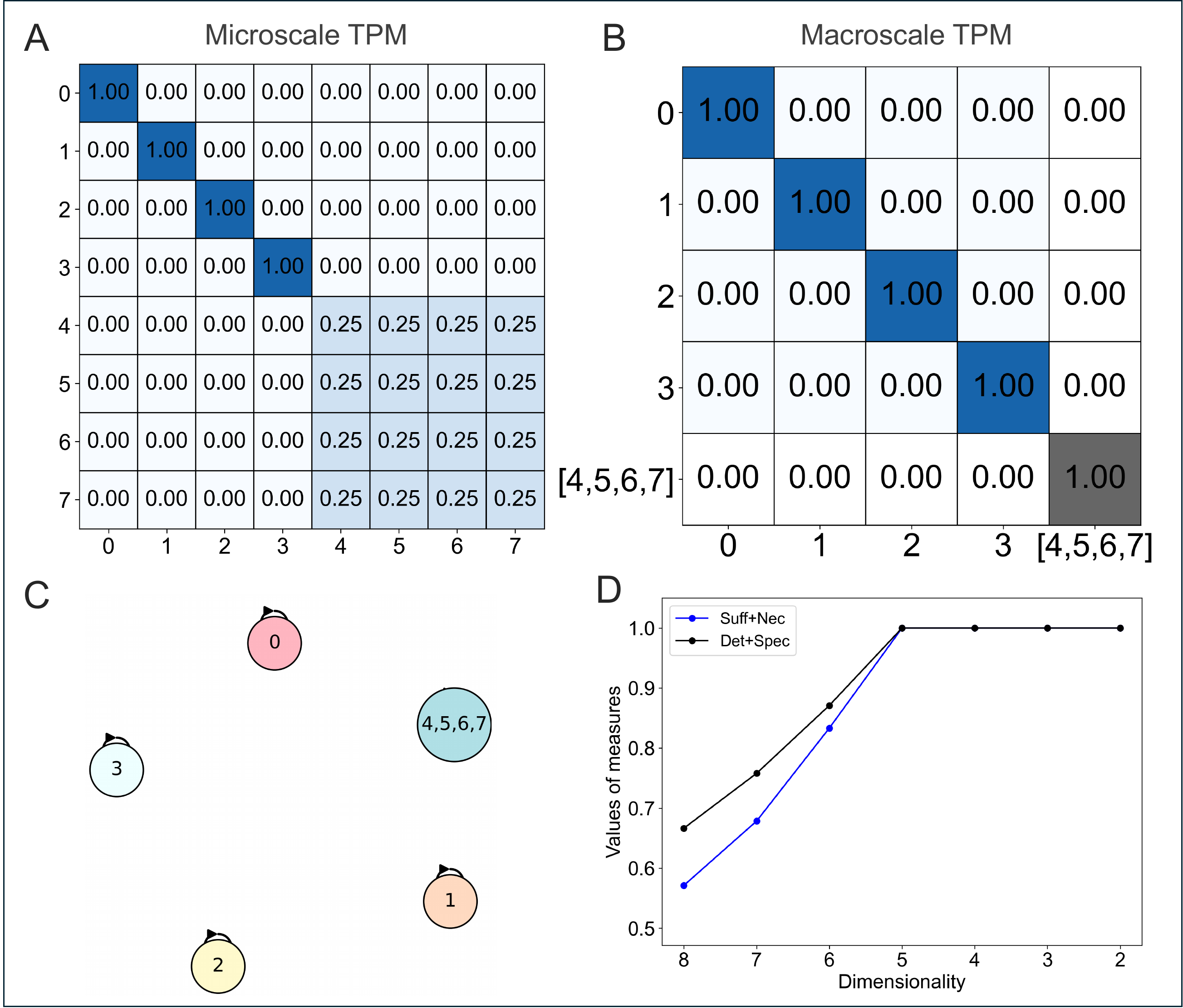} 
    \caption{\textbf{Causal primitives along a micro$\rightarrow$macro path}. (A) The TPM of the microscale, with cells colored based on their probability ($p=1$ being a darker blue). (B) The TPM of the macroscale past which \(\Delta \mathrm{CP}\) transitions abruptly to zero. (C) The same macroscale visualized as a Markov chain, with the coarse-grained macrostate labeled (its self-loop of $p=1$ is not shown). (D) The change in causal primitives across the path of increasing dimension reduction, with the total gain in CP of $0.33$ (in terms of determinism plus specificity) reflecting the degree of causal emergence.}
    \label{fig:4}

\end{figure}

Notably, in this system CP shows consistent gains until the intuitive macroscale of $(4, 5, 6, 7)$ are coarse-grained together into a single macrostate with a self-loop (the TPM of this macroscale is shown in Fig.2B, and is visualized in Fig.2C). Then the path immediately transitions to a domain of zero gains (plotted in Fig.2D).

Expressed in terms of determinism plus specificity, the microscale starts at CP $=0.66$, and at the macroscale prior to the transition to zero gains this has increased by $0.33$, attaining a maximum value of CP$=1$, indicating that causal relationships become maximally deterministic and non-degenerate at that macroscale, and further coarse-graining entails no further gains. The degree of causal emergence in this system is therefore CE $= 0.33$, reflecting the total gain along the path.

\subsection{Choosing micro$\rightarrow$macro paths}

When analyzing the causal emergence of a system, there may be prior reasons to pick a certain path; however, an appropriate micro$\rightarrow$macro path can also be identified in a first-principles manner.

Specifically, the endpoint of the micro$\rightarrow$macro path can be chosen as the macroscale which entails the highest total gain in CP possible (the maximum causal emergence). In the case where there are multiple possible endpoints tied for the highest gain, the macroscale representing the lowest amount of dimensionality reduction is the optimal endpoint, as it indicates the point past which dimensionality reduction does not lead to gains in CP. Once the endpoint is identified, the maximally-informative micro$\rightarrow$macro path to analyze \(\Delta \mathrm{CP}\) along is the longest path across the set of consistent macroscales (with consistency defined as in S2 in the SI) ranging from the microscale to the endpoint macroscale.

The macroscale which serves as the endpoint can be identified in a brute-force search by first generating all possible macroscales, discarding those which are inconsistent, calculating their CP, and then choosing the macroscale with the highest CP and highest dimensionality as the endpoint to the path.

In practice, this is not feasible for larger systems, and heuristics are required. One is to coarse-grain along a path until diminishing returns are reached. To put this formally, there is a sequence of incremental gains \(\Delta \mathrm{CP}_i\) 
at each step \(i\) in the path. The system enters ``diminishing returns'' at step \(i^*\) 
if \(\Delta \mathrm{CP}_{i^*} < \varepsilon\) (for some small threshold 
\(\varepsilon > 0\)), or if the ratio 
\(\Delta \mathrm{CP}_{i+1}\!/\!\Delta \mathrm{CP}_i\) consistently decreases 
over a long path length.

In other words, once \(\Delta \mathrm{CP}\) becomes negligible (below \(\varepsilon\)) or keeps shrinking step-by-step for some long portion of the path, that indicates a transition point to diminishing returns and an approximate endpoint that does not leave out substantial gains in CP. However, care must be taken to not simply stop at local maxima (see Section 4) by choosing a small enough \(\varepsilon\) or a long enough path length to assess diminishing returns. 

For a bound of the amount of causal emergence in a system prior to defining a path or even any macroscales, one can simply measure CP at the microscale. Its distance from 1 provides an upper bound for CE without the need to search across scales, allowing for quick estimations. If the CP of the microscale is significantly less than 1, it is likely that there is some dimension reduction (like a coarse-grain) that increases it to at or near maximum; therefore, the difference between the CP of the microscale and 1 may approximate the CE value for many systems (but does not specify which macroscale the gain comes from). It is even possible that for large enough systems with enough structure there is always some macroscale where CP approaches 1, especially when taking into account the full set of dimension reductions, like Higher Order Macrostates \cite{klein2020emergence}, or when relaxing consistency assumptions.

\section{The emergent complexity}
\label{sec:multiscale}

Traditionally, a main motivation of the field of complexity science is the qualitative notion that there is intuitive macroscale or mesoscale structure to the system beyond the microscale. Specific quantitative methods for detecting macroscale or mesoscale structure have focused mainly on compressibility or efficiency \cite{gershenson2012complexity}, or more recently, by assessing the info-theoretic surprise \cite{marchese2022detecting}. However, this means that the mesoscales detected may just be convenient compressions and have no causal relevance. 

In comparison, by analyzing the distribution of causal contributions, CE 2.0 can be used to quantify the emergent complexity that's actually causally relevant for the workings of the system. Specifically, the \(\Delta \mathrm{CP}\) for each step along a path represents that scale's causal contribution to the total CP (which fully determines the system's causal workings)---and so its distribution along a path can be assessed. CE 2.0 therefore provides a taxonomy for how complex the causal workings of a system are: if they are mostly confined to a single scale (like either the microscale, or in that it is dominated by a “top-heavy” macroscale) then the system is simple, whereas if the system has intermediate mesoscales that also have substantial causal contributions, it is complex.

\begin{figure}[H]
    \centering
    \includegraphics[width=1\textwidth, height=0.8\textheight, keepaspectratio]{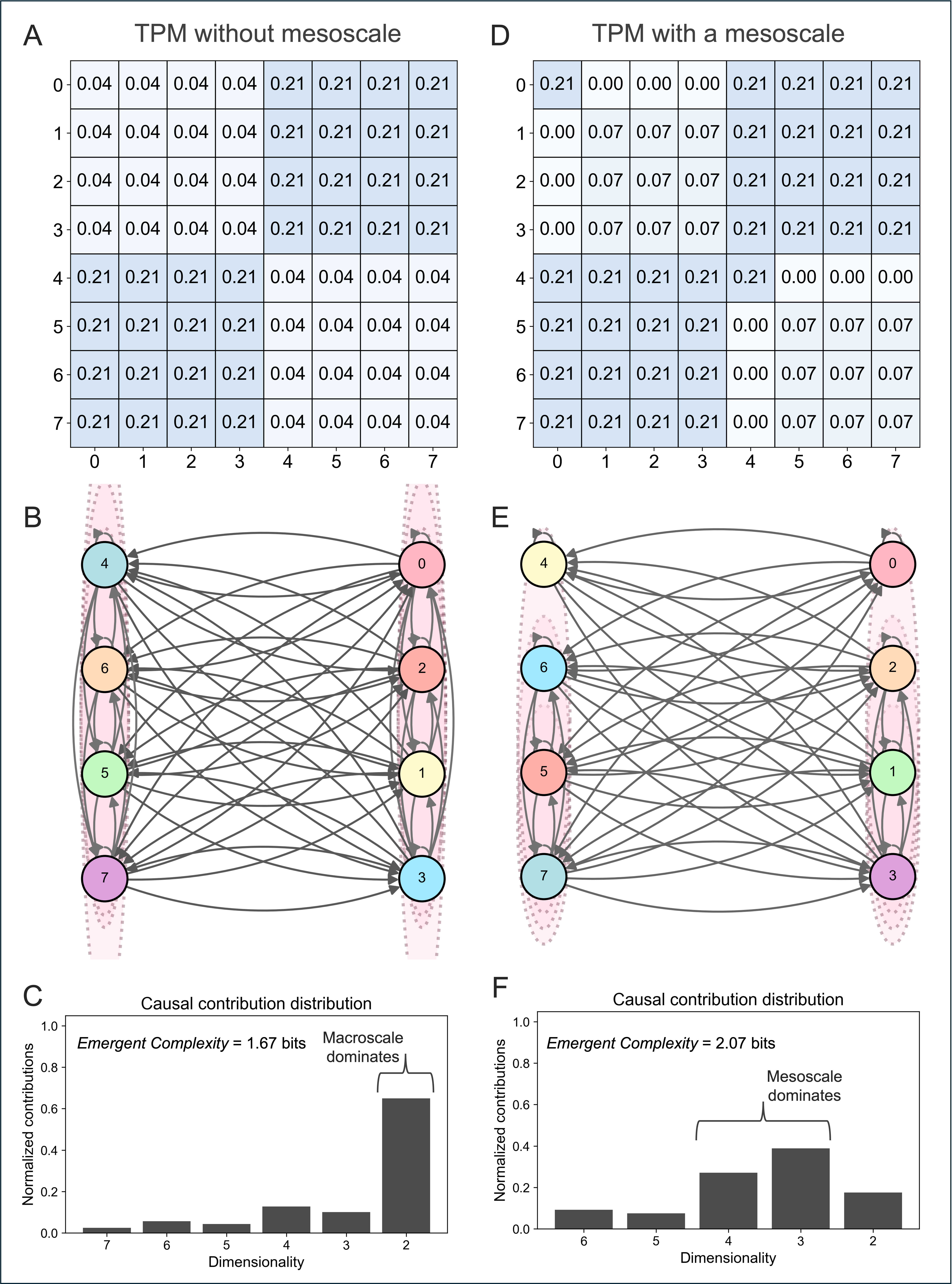}
    \caption{\textbf{Causal contributions across scales}. (A) Microscale TPM of a system with no mesoscale structure. (B) The same system visualized. (C) CE 2.0 identifies this system's causal contributions as ``top-heavy,'' in that the last dimension reduction contributes the most. (D) Microscale TPM of an otherwise similar system with mesoscale structure. (E) The mesoscale system visualized. (F) Causal contributions are shifted toward lesser dimension reductions, indicating a predominately multiscale causal structure; it therefore possesses more emergent complexity.}
    \label{fig:5}

\end{figure}

To demonstrate this novel aspect of CE 2.0, in Figure~\ref{fig:5} two causally emergent systems are analyzed: one that has no mesoscale structure and one that does, but is otherwise as similar as possible. 

Fig.3A shows the TPM of a system composed of an intuitive macroscale with no distinguishable mesoscale structure, wherein $(0,1,2,3)$ and $(4,5,6,7)$ have been coarse-grained into two respective macrostates. Indeed, the microstates coarse-grained into the two macrostates each make up an equivalency class (see Fig.3B for a visualization). 

At each scale along the micro$\rightarrow$macro path, \(\Delta \mathrm{CP}\) is tracked. Notably, for the first system, gains in CP along the micro$\rightarrow$macro path are clustered at the endpoint of the path (see Fig.3C). This indicates a ``top-heavy'' structure mostly composed of the causal contributions of the microscale (contributing 0.14 of the total CP) and the endpoint of a large macroscale over the two equivalency classes (contributing 0.18). Together the microscale and macroscale account for the majority of the full CP value (.41) for the system, quantifying that its causal workings are dominated by those two scales.

To formally capture the taxonomy of ``top-heavy'' or ``bottom-heavy'' systems on one side, and systems with substantial mesoscale structure on the other side, here I introduce a notion of \textit{emergent complexity} (EC). It is based on the entropy of the causal contributions along a path of length $L$. Given a set of gains $\Delta \mathrm{CP}_i$ at each step (excluding the microscale) $i=1,2,\ldots,L$, then
\[
p_i \;=\; \frac{\Delta \mathrm{CP}_i}{\sum_{j=1}^{L} \Delta \mathrm{CP}_j}
\quad
\text{for } i = 1,\dots,L
\]
which ensures that $\{p_1,\dots,p_L\}$ is a probability distribution over the $L$ steps. To measure how ``spread out'' (multiscale) the gains are, the entropy is calculated
\[
\text{EC} = -\sum_{i=1}^{L} p_i \log_2(p_i)
\]
which equals $\log_{2}(L)$ if all $\Delta \mathrm{CP}_i$ are equal (i.e., $p$ is uniform), and decreases when a small number of steps in the path dominate the total gains, reaching zero when a single emergent scale has a lone causal contribution. To compare paths of significantly different lengths, these values in turn can be normalized by $\log_{2}(L)$.

To demonstrate how this method detects mesoscale structure, a similar system is modeled in Fig.3D-E, but with the change that $(0)$ and $(4)$ are distinguishable (in terms of their state transitions) from the other of members they are coarse-grained together with at their endpoint macrostate, $(0,1,2,3)$ and $(4,5,6,7)$, respectively. That is, for this system the largest dimension reduction that contributes gains to CP is not over a pure equivalency class.

In this mesoscale system (with a total CE of 0.13 at the endpoint where $(0,1,2,3)$ and $(4,5,6,7)$ are macrostates), there is a visible earlier maxima when the causal contributions along the path are plotted (Fig.3F). Because of this earlier maximum in terms of \(\Delta \mathrm{CP}\), the emergent complexity of the mesoscale system is 2.07 bits, whereas the emergent complexity of the ``top-heavy'' system is only 1.67 bits. That is, mesoscales are revealed by how \(\Delta \mathrm{CP}\) peaks at a higher dimensionality than the endpoint.

\section{Limitations and heuristics}

A limitation of CE 2.0 is that, in its current formulation, it assumes a particular chosen micro$\rightarrow$macro path to define a hierarchy of scales. This initial formulation makes practical and conceptual sense; especially since a single path might often be desired for analysis. However, work remains in developing a causal apportioning schema that can integrate the full set of non-commensurate micro$\rightarrow$macro paths; potential opportunities are in the development of a causal apportioning schema that operates across the full set of partitions, making use of tools like the Möbius inversion \cite{jansma2025mereological} or the Shapley values \cite{winter2002shapley}.

A more practical limitation for CE 2.0 is that traversing the set of macroscales of a system entails a combinatorial explosion. Previous heuristics exist for CE 1.0 \cite{zhang2022neural, griebenow2019finding} (and in continuous systems \cite{chvykov2020causal}), including most recently the work by Zhang et al. \cite{zhang2025dynamical}, which proposed using the singular value decomposition (SVD) as a precise way to estimate the gain in EI available at some macroscale without having to search across the full set of scales (and thus avoiding the associated combinatorial explosion). This work also highlighted the importance of dynamical reversibility to causal emergence, in that their measure of dynamical reversibility behaved very similarly to the determinism and degeneracy (indicating that the causal primitives have a joint relationship to reversibility).

Therefore, in the Supplementary Information herein a novel heuristic for CE 2.0 is explored, based on adapting their SVD method. As detailed in S3, this method can detect the differences in multiscale structure of the systems compared in Section 4 without requiring a search across macroscales, indicating that the CE 2.0 analysis, such as the causal emergence and the emergent complexity, can be estimated without combinatorial explosions.

\section{Comparison to other theories of emergence}
\label{sec:comparison}

CE 2.0 has many advantages as a theory: (a) it is axiomatically grounded in the most foundational terms of causal analysis and is robust to the assumptions within that analysis, (b) it captures all possible cases of macroscale causation, which CE 1.0 does not, and (c) it elucidates the multiscale causal structure of systems in a novel manner, resolving longstanding conflicts around over-determination and causal exclusion. Here, these advantages are considered and demonstrated.

\subsection{An axiomatic grounding}
\label{sec:axiomatic_grounding}

As discussed, the EI's use in CE 1.0 has been previously criticized on the basis of being drawn from a maximum entropy (uniform) distribution \cite{balduzzi2011information} (here, represented by some $P(C)$) \cite{aaronson_higher-level_2017, eberhardt2022causal}. The critical issue was that CE 1.0 required this assumption of the EI to detect causal emergence---and indeed, a theory of causal emergence should not be reliant on the assumptions behind the EI calculation. This would impact both its practical uses and also its theoretical foundations. In some cases, like in Integrated Information Theory, the maximum entropy distribution can be further justified as a function of taking the ``intrinsic perspective'' on a system \cite{marshall2024micro}, but this is reliant on accepting the postulates of IIT, including its analysis of consciousness.

In CE 2.0 it is also recommended that a uniform distribution is used for $P(C)$. This is because not allowing $P(C)$ to differ between scales would imply counterfactuals and interventions at a macroscale cannot be calculated independently from their microscale; a nonsensical outcome for most causal models of science (e.g., assessing the strength of the causal relationship between a light switch and a light bulb would entail weighting by the total atoms in each, etc.). 

However, unlike in CE 1.0, this recommendation for $P(C)$ is not required to detect causal emergence. Indeed, gains in CP at the macroscale have been shown to be robust to choice of $P(C)$, to the degree of gains even occurring when using the observed distribution at both the microscale and macroscale \cite{comolatti2022causal}. This advantage can also be seen directly in Figure 4, wherein the $P(C)$ of both the microscale and the macroscale are the observed distributions at the respective scales, and thus the macroscale's $P(C)$ is itself directly a coarse-grain of the $P(C)$ of the microscale (meaning any description of intervention distributions are equivalent between the scales). Yet CE 2.0 still detects macroscale causation under those conditions.

In sum, CE 2.0 is more theoretically robust than CE 1.0 was, by virtue of being grounded in causal primitives that historically have shown to be fundamental to the nature of causation and by detecting causal emergence across diverse background assumptions in how the causal analysis is performed.

\subsection{CE 2.0 captures all macroscale causation}
\label{sec:all_macroscale_causation}

CE 2.0 can detect cases of macroscale causation that the CE 1.0 framework does not.

Examples of this detection are shown in an 8-state system that is constructed of two equivalency classes, making it a ``block model'' at the microscale (see Fig. 4A, left). A single macroscale is specified, wherein the two equivalency classes are each coarse-grained into a respective macrostate with a self-loop, represented by the coarse-grain $(0,1,2,3), (4,5,6,7)$. Causal emergence as calculated via CE 1.0 (based on the gain in EI at the macroscale) and the causal emergence as calculated via CE 2.0 (based on the gain in CP at the macroscale) is shown across a manipulation of that system. 

Starting with the initial TPM shown in Fig.4A (left), the probabilities within each equivalency class for each state $s_i$ are manipulated such that, over 50 steps, the probabilities that were previously transitioning to the other members of the equivalency class are added incrementally to the self-loop probability of $s_i$, eventually reaching $p=1$. The midpoint of this manipulation is shown in Fig.4A (middle), and the final ending system is the TPM in Fig.4A (right). This progresses the system via discrete steps of probability redistribution from an initial ``block model'' configuration to a permutation matrix in the form of a set of 8 microstates with self-loops of $p=1$. 

At each step calculations from CE 1.0 are shown based on the EI, compared to the gain in CP as in CE 2.0 (with an endpoint of the fixed chosen macroscale). Counterintuitively, the EI from CE 1.0 detects no causal emergence, even when the system is initially split into two equivalency classes. Meanwhile, when viewed from the new lens of CE 2.0, the system in Figure 4 sensibly starts with a significant degree of macroscale causation. This degree then decreases in accordance with the increasing self-loop probabilities and the increasing distinguishability of the microscale, becoming weaker and weaker as the macroscale contributes marginally less, until it vanishes altogether when the microscale becomes perfectly deterministic and non-degenerate. 

\begin{figure}[H]
    \centering
    \includegraphics[width=1\textwidth]{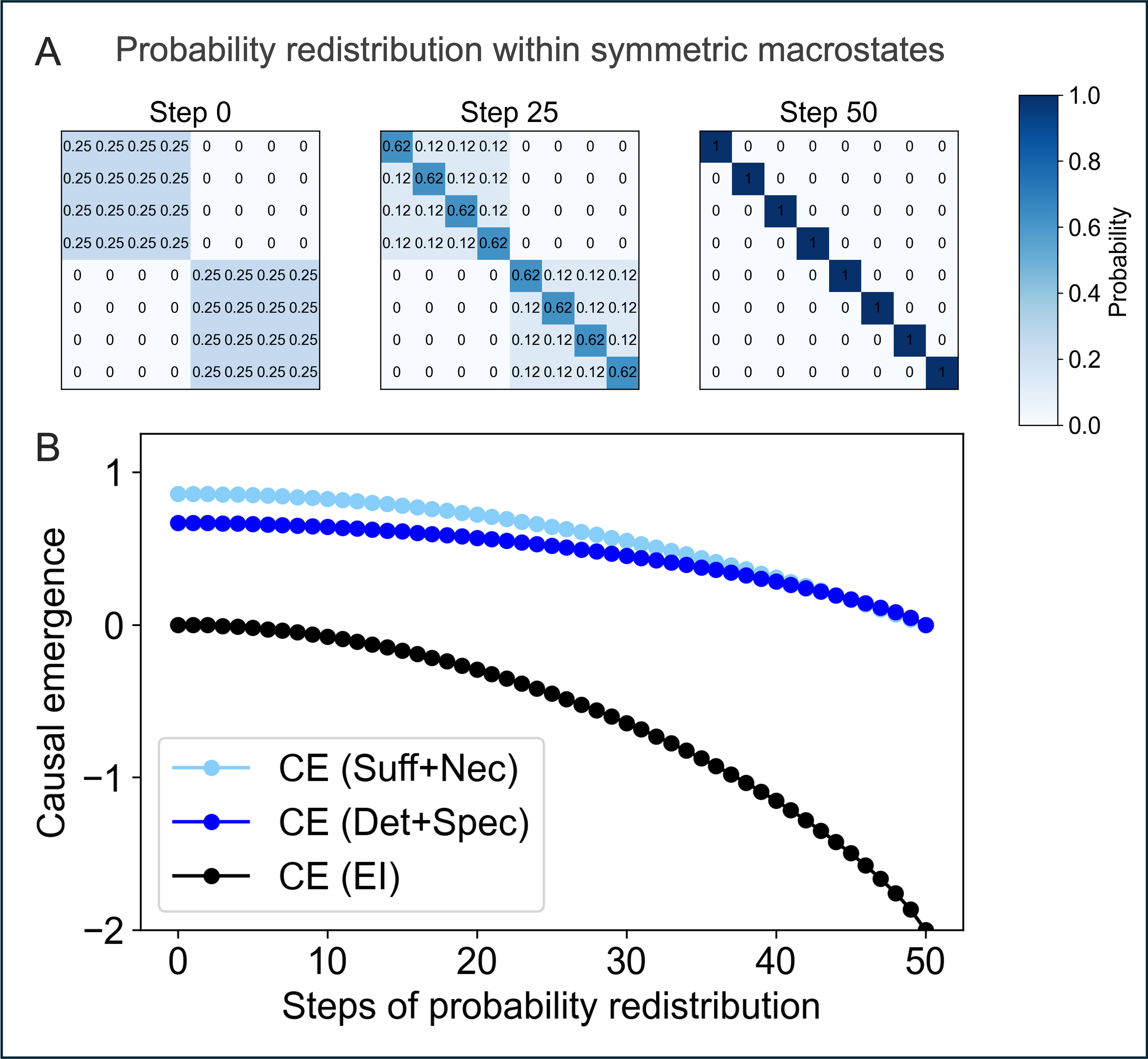} 
    \caption{\textbf{CE 1.0 cannot capture all macroscale causation.} (A) TPMs (probabilities shown in bluescale) of a ``block model'' system with two macrostates over its equivalency classes at the beginning, midpoint, and end of increasing the self-loop probabilities of each state. Redistribution is performed by drawing away probability from its full set of transitions via increments of $1/steps$ until the microscale is entirely composed of states with self-loops of $p=1$. (B) CE 2.0 detects the macroscale causation and decreases sensibly as the microscale becomes more causally distinguishable during the probability redistribution, while the EI does not.}
    \label{fig:6}

\end{figure}

In some cases, CE 1.0 and CE 2.0 will overlap (e.g., the scale right before \(\Delta \mathrm{CP}\) transitions to being zero in Figure 4 is the same as would be picked out by searching for the maximum of EI). This is because CE 1.0 and CE 2.0 share a close mathematical connection in their terms, since the EI has a decomposition wherein:

$$ \textit{EI} = \textit{effectiveness}  * {\log_2 n} $$

That is, the EI can be decomposed into the determinism minus the degeneracy (the \textit{effectiveness}), which is then multiplied by a \textit{size} term, ${\log_2 n}$, which in turn is just the dimensionality of the given scale (its number of states) \cite{hoel2013quantifying}. However, in CE 2.0, the \textit{size} term is rendered unnecessary by appropriate multiscale structure analysis and causal apportioning.

While CE 2.0 focuses on \(\Delta \mathrm{CP}\), if a single causally-relevant scale with high dimensionality is desired for causal modeling or explanation, its methods can be used to pick one out: the individual macroscale that maximizes CP with the smallest amount of dimensionality reduction. This can be done either by analyzing for diminishing returns in \(\Delta \mathrm{CP}\), as described in Section 3.3, or even via explicit re-introduction of the \textit{size} term against which to weigh gains in CP, flexibly re-capturing the analysis of CE 1.0.

\subsection{Comparison to other related theories of emergence}
\label{sec:related_theories}

Alternative proposals have been suggested for different ways to detect causal emergence, or emergence more generally, such as via integrated information decomposition \cite{rosas2020reconciling}, or via examining dynamical dependency \cite{barnett2023dynamical}. However, both these proposals make use of the mutual information. This is problematic for causal emergence, as it is definitional of causation that it is not dependent on simply the data distribution of the process being measured \cite{pearl2018book}. For example, in a cycle of COPY gates, the mutual information scales entirely off of how varied the initial state is, rather than capturing the fact that every gate is sufficient and necessary for the next step in the chain in the way the causal primitives do \cite{comolatti2022causal, zhang2025dynamical}.

Other recent work on emergence has focused on identifying cases wherein macroscales are consistent with their underlying microscales, but still independently describable in their dynamics, like the software of a computer \cite{rosas2024software}. Such efforts have examined whether or not a macroscale is “causally closed” in that it can can be thought of as being its own cause. This is closely related to the condition of macroscale consistency based on random walkers laid out here and (to a less strict degree) previously as well \cite{klein2020emergence}. However, merely checking for consistency, lumpability, causal closure, etc, does not directly measure causal emergence, as it does not reflect what a macroscale contributes to a system's causal workings above and beyond the microscale, which requires some specified measure of causation (or here, gains in the causal primitives that underly such measures). Rather, it merely identifies which macroscales are valid descriptions of their microscale that preserve its dynamics and therefore are appropriate compressions. E.g. for the system analyzed in Figure 4, the largest dimension reduction at the end the pre-chosen path is a valid macroscale, completely consistent with its microscale, and yet has trivially zero causal contribution.

\subsection{Conceptual implications of CE 2.0}
\label{sec:philosophy}

All of science, outside of microphysics, implicitly operates as if there is causal emergence, in that it takes for granted the macroscale entities in its models and explanations and experiments are efficacious regarding a system's causal workings. This is contradicted by a nominal commitment to universal reductionism, which would seem to imply that all causal powers “drain away” to the bottom microscale of any system \cite{bontly2002supervenience, block2003causal}. This is an effect of the causal exclusion argument \cite{kim2000mind}---for any given supervening macroscale, its effect could also be described as a cause of its underlying microscale, which then renders the macroscale description unnecessary.

CE 1.0 flipped the exclusion argument on its head by noting that, according to the EI, the macroscale had greater causal power. A similar sort of thinking underlies the exclusion postulate in Integrated Information Theory (which is perhaps the most controversial of the postulates \cite{bayne2018axiomatic}). But it meant that in CE 1.0 there was the counterintuitive result that, even when macrostates were not over exact equivalency classes, the underlying microscale could itself be excluded; a surprising result with unclear epistemological and ontological implications.

Comparatively, in CE 2.0 causal exclusion is handled more gracefully. When viewed from the current analysis of a single path, macroscales do not override the causation of the microscale (although the microscale's causal contribution can still be vanishingly small). Instead, they simply contribute additional causal power via the causal apportioning schema, leading to a more comprehensive mereology wherein individual scales are a lossy slice of a higher-dimensional object that contains all the relevant information about the system's causal workings. 

While other theories of emergence often posit that certain macroscale properties or laws are not even in-principle reducible to the microscale, and therefore physics is not “causally closed” (e.g., \cite{anderson1972more, ellis2020causal, fritzman2024collapsing}), this remains a controversial requirement for theories of emergence \cite{carroll2017big, carroll2024emergence}. Comparatively, causal emergence can occur even when macroscales are fully reducible to their microscales (like the models herein), as macroscales reduce uncertainty in causal relationships regardless. That is, even when macroscales themselves are reducible, their gains in causal primitives are by definition not. The source of these gains beyond the microscale is non-mysterious, being based in uncertainty reduction \cite{hoel2017map, liu2024exact}, which in turn stems from the multiple-realizability of macrostates \cite{hoel2024world}.

Since in CE 2.0 emergence occurs via this minimization of uncertainty at macroscales, there is the broader question of whether uncertainty (in the form of noise or common causes) exists merely epistemically when it comes to the causal models of science and the systems they represent. Answering this involves speculation about unknowns like the scientific end-state of physics \cite{hoel2024world}. Even small sources of true uncertainty (like indeterminism) can be amplified in chaotic systems, and there may even be provably undecidable physical systems \cite{cardona2021constructing}. Even if all uncertainty inherent to the causal relationships of scientific causal models did turn out to be only in principle epistemic, this would hold only for closed causal models that span the entire universe, and in such a universe-size causal model all notion of causation disappears entirely anyway, as there are no definable interventions from outside the model \cite{pearl2009causality}. So while causal emergence vanishes in the condition of a microscale possessing zero uncertainty about the effect of causes, and also possessing no common causes (e.g., like a permutation matrix), such conditions entail a far departure from most causal models in science.

\section{Conclusion}

CE 2.0 provides a conceptually and mathematically novel theory of emergence that treats systems as a hierarchy of scales. Individual scales, even in most cases the microscale, are simply slices of a higher-dimensional object---but only a slim minority of these scales are causally relevant, and the theory can identify them, revealing the hierarchy that matters to a system's causal workings. 

Specifically, the theory shows how macroscale causation is measurable via the gain in information-theoretic generalizations of the causal primitives (sufficiency and necessity) along a specified path that traverses the possible dimension reductions of a system. Causal emergence is the sum of this gain, and causal contributions can be apportioned out along the path. The theory reveals a novel taxonomy wherein some systems are “top-heavy” in terms of their causation, like when a single macroscale dominates, whereas others have more mesoscale structure. The emergent complexity calculation quantifies how widely spread out across the hierarchy of scales causal contributions are. 

The theory's initial formulation, based on a single path that traverses dimension reductions, does have some limitations. It does not specify how to apportion across multiple paths, and it faces combinatorial explosions in application. However, how to address these limitations is relatively clear (see Section 5 for a discussion of pathless approaches, and see S3 for the proposal of a heuristic for CE 2.0's analysis).

CE 2.0 has critical applications in fields like physics, biology, neuroscience, and economics. Here, it's also worth highlighting one new particular use case: Given the axiomatic importance of causal primitives for understanding causation in complex systems, and given previous research showing the EI is responsive to changes following learning in artificial neural networks \cite{marrow2020examining}, CE 2.0 may also be primed to contribute to the growing field of AI interpretability \cite{templeton2024scaling} and AI safety \cite{lazar2023ai}, such as analyzing the multiscale structure of deep neural networks.

\section{Acknowledgments}
\label{sec:acknowledge}

Thank you to Softmax for funding the development of CE 2.0 and the writing of this paper. Thanks as well to Adam Goldstein and Emmett Shear for valuable feedback on the draft, and also to Michael Levin for encouragement and conversations.

\clearpage
\appendix
\renewcommand\thefigure{S\arabic{figure}}
\renewcommand\thetable{S\arabic{table}}
\renewcommand\thesection{S\arabic{section}}
\setcounter{figure}{0}
\setcounter{table}{0}
\setcounter{section}{0}

\section*{Supplementary Information}
\addcontentsline{toc}{section}{Supplementary Information} 

\section{Causal primitives (and their generalizations) are sensitive to noise and common causes.}

Previous research on causal consilience has already shown that, due to their close mathematical relationship, the sufficiency and necessity over a set of two transitions behaves similarly to their information-theoretic generations, determinism and degeneracy, in conditions of increasing uncertainty \cite{comolatti2022causal}. Here, their similarity is shown with a larger set of transitions.

Specifically, a system composed of 8 states with self-loops of $p=1$ was specified (see Fig.S1, top left). In order to vary the uncertainty about causes and effects, noise (uncertainty about effects) and common causes (uncertainty about causes), were introduced along two separate axes. The first axis increased uncertainty about effects by shifting the system down to a condition of complete randomness in terms of its transitions (an all-to-all Markov chain wherein all transitions = $1/n$, which means the system behaves as unpredictably as possible). The second axis, uncertainty about causes, moved the system to the condition wherein all transitions had an identical set of effects (thus increasing the number of common causes). 

For every step along the axis that increased the noise, the probability from each self-loop was redistributed equally across the other states in the system, with the total amount of probability redistributed being $1/steps$ each step. For every step along the axis that increased the number of common causes, the full set of transitions for a state (a row in the TPM) were replaced one at a time with a duplication of the first row until all distributions were the same. Changing the model along just this latter axis began with all states having unique state-transitions and ended with all states transitioning to a single state (see Fig.S1, bottom left). Finally, these two axes of changes to the system were combined such that at every step, both more noise in effects was introduced, and at the same step, more common causes were introduced (see Fig.S1, middle diagonal).

At each step, the system-wide sufficiency plus the necessity was calculated for each state, as well as the determinism plus the specificity, along the increasing uncertainty in effects axis (Fig.S2A), along the increasing common causes axis (Fig.S2B), as well as along both axes combined (Fig.S2C). As in the main text, this is done in a way to ensure the same $[0, 1]$ bounds (See Section 3.2).

\begin{figure}[H]
    \centering
    \includegraphics[width=1\textwidth]{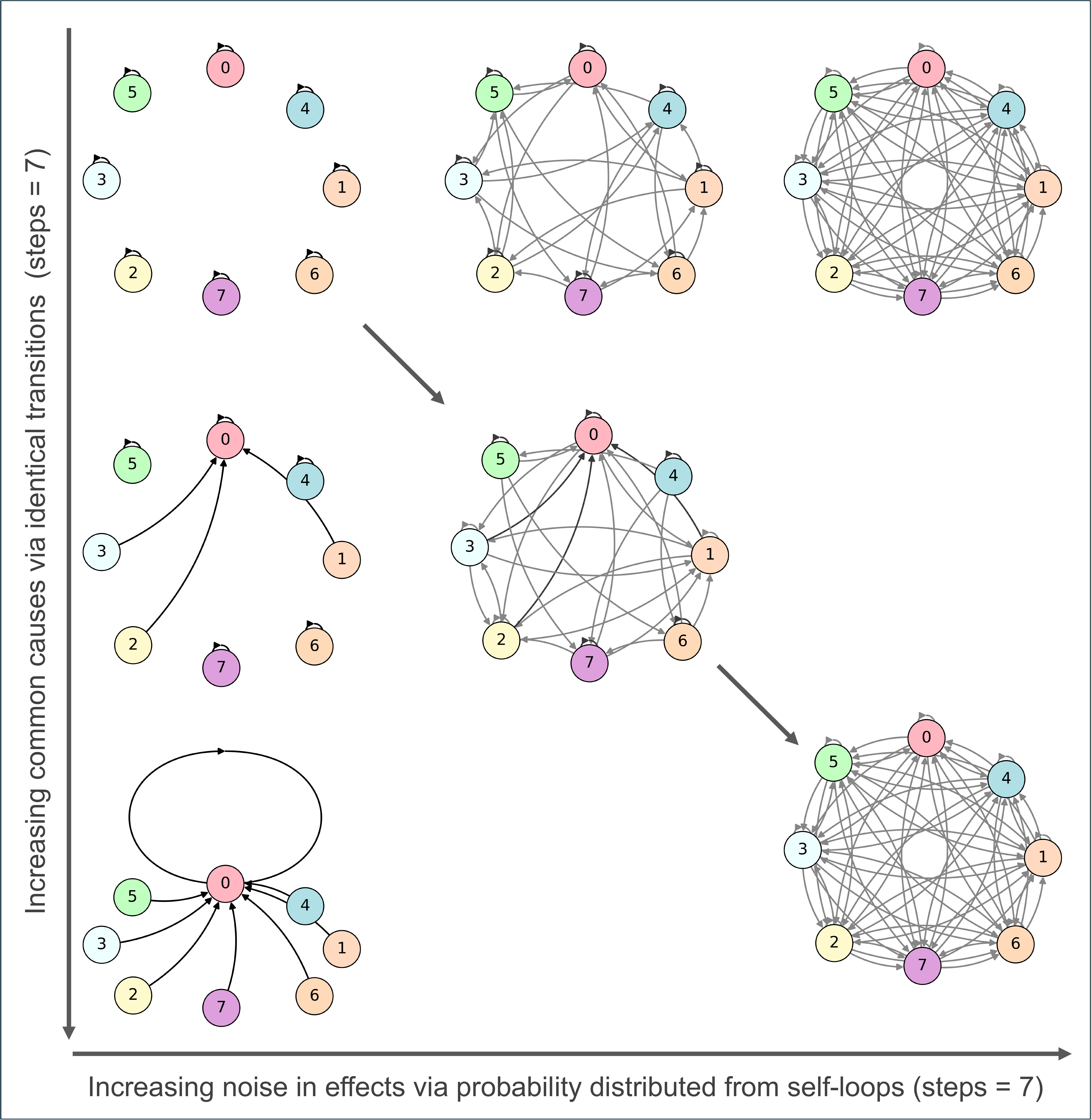} 
    \caption{\textbf{Increasing uncertainty in the causal relationships of an 8-state system}. Starting in a state of self-loops with $p=1$, states in the network were changed over a set number of steps equal to the size (number of states) of the system in three ways. Along the x-axis, self-loop probabilities were reduced by $1/steps$ and distributed equally to the other states (thus increasing the uncertainty of a particular effect, given a cause) until the system was an all-to-all network with random transitions. Along the y-axis, at each step a state was replaced with the transition distributions of another state (increasing the number of common causes and thus increasing the uncertainty of a cause, given an effect), until all states in the system shared the same transition. The system was also subjected to both changes at each step (the middle diagonal), ending again in an all-to-all state of random transitions.}
    \label{fig:1}

\end{figure}

Notably, all CP values changed similarly along both axes of steps that increased uncertainty about causes and effects, even in larger systems (values for 100 states and 100 steps are also shown in Fig. 2D).

\begin{figure}[H]

    \centering
    \includegraphics[width=1\textwidth]{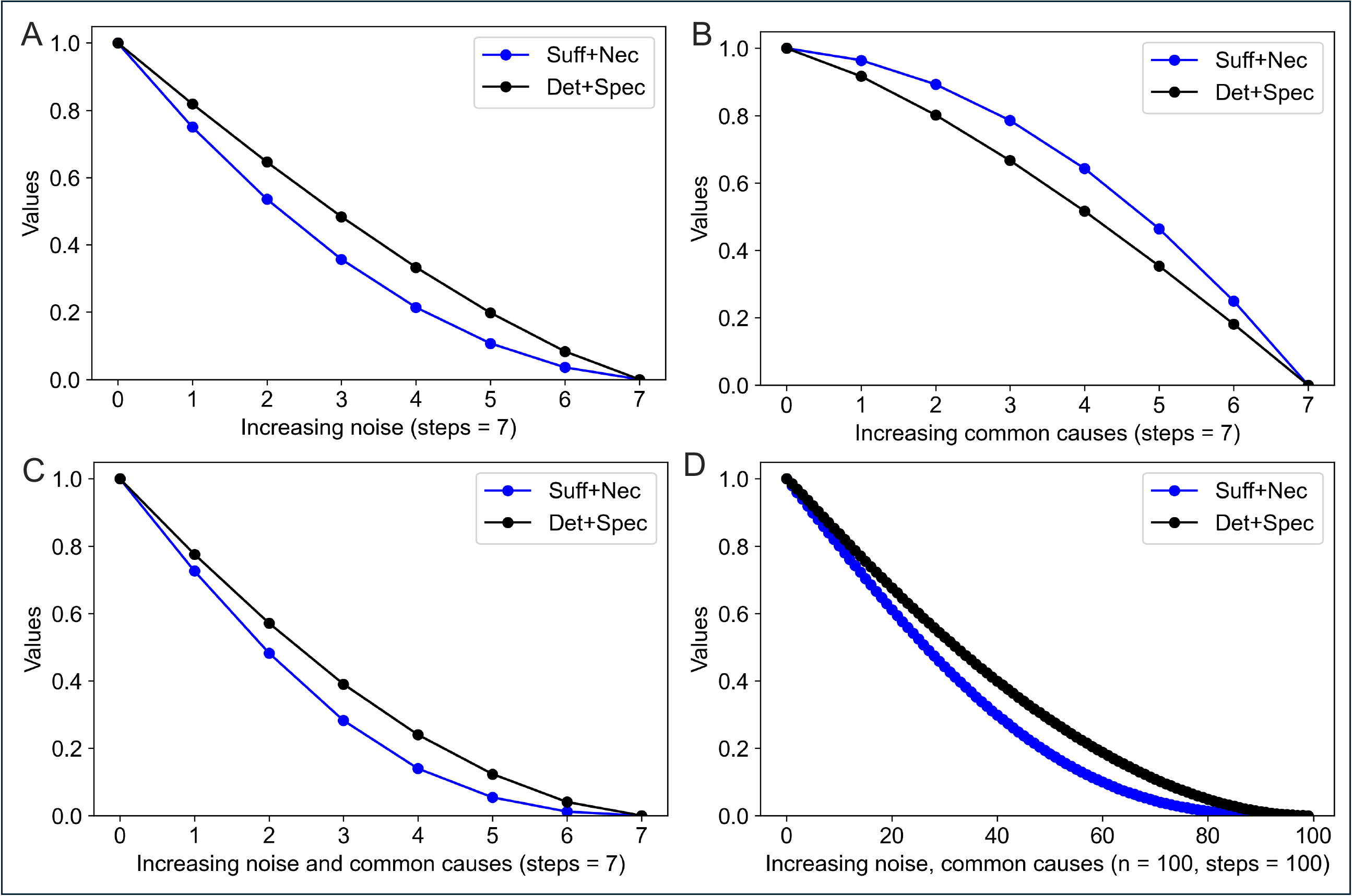} 
    \caption{\textbf{Causal primitives vary together with uncertainty.} (A) The sufficiency plus necessity value and the determinism plus specificity value are shown to behave similarly across increasing uncertainty over effects (noise) as the probabilities of self-loops in the system (visualized in Figure S1) are redistributed across the system in increasing steps. (B) Similar behavior in response to increases in common causes (overlap). (C) Similar behavior in response to increases in both noise and overlap. (D) The measures still behave similarly as the system becomes larger and more steps are added.}
    \label{fig:2}

\end{figure}

\section{Calculating dynamical consistency across scales.}

Not all macroscales are sensible summaries of their underlying microscale; indeed, some macroscales may be dynamically inconsistent when defined \cite{rubenstein2017causal}. Much as in Klein and Hoel \cite{klein2020emergence}, here I deem a macroscale valid if it is \textit{consistent} with its underlying microscale, with consistency being defined as whether the path of random walkers on the Markov chain is the same at both the microscale and the macroscale (i.e., whether or not the macroscale acts as an accurate summary statistic for the microscale's dynamics).

Specifically, for a Markov chain the inconsistency can be defined as the Kullback-Leibler divergence \cite{cover1999elements}, taken between an expected distribution of random walkers, across timesteps $t$ $\rightarrow$ $t_n$, in $S$ vs. $S_M$, given an identical starting state on each scale. While previous research checked only the stationary distributions for such inconsistency \cite{klein2020emergence}, here I enforce a strict notion of consistency, wherein a random walker is dropped at every possible state, and inconsistency between the macroscale and the microscale is summed over all of their moves for the next 5 timesteps. Any non-zero values imply inconsistent macroscales, which are discarded. Therefore, all macroscales considered herein are fully consistent with the dynamics of their underlying microscales.

\section{Heuristics for the CE 2.0 framework via the SVD}

Zhang et al. \cite{zhang2025dynamical} proposed a “vague” causal emergence calculation based on applying the singular value decomposition (SVD) to Markov chains. This is based in the average of the resultant singular values ($\sigma$s), which they call $\gamma$, and which they prove reflects the average dynamical reversibility across system states. It was also shown capable of approximating the determinism plus specificity used herein (for details, see \cite{zhang2025dynamical}). As they noted, this indicates a connection between the causal primitives and dynamical reversibility.

Furthermore, their research provided a way to approximate (quite precisely) the degree of causal emergence in the CE 1.0 framework (analogous to finding the macroscale with maximal EI). Specifically, they first identify a set of non-zero $\sigma$ values, given some threshold, $\epsilon$, and average these together ($\bar{\sigma}$). The difference between $\gamma$ (the un-thresholded average) and $\bar{\sigma}$ is taken. When $\epsilon$ is a small non-zero value, and so includes most non-zero $\sigma$ values, this method can elegantly track the maximal increase available in EI at some possible macroscale, without have to search across the set of scales, and so offers a precise heuristic for the degree of causal emergence in the CE 1.0 framework without any combinatorial explosions (only requiring the TPM of the microscale).

Here I show steps for how the SVD method can be adapted for the CE 2.0 framework (in which case, “vague” causal emergence becomes a specific value). To do this, the set of singular values can be used as proxies for \textit{directionalities} of coarse-graining. The causal contributions of each directionality (represented by each $\sigma$) can then be calculated via an adaptation of the causal apportioning schemes detailed in Section 3.2.

Here, unlike in \cite{zhang2025dynamical}, the initial trivial $\sigma_1$ value is discarded, since it is always 1 or greater for any Markov chain and therefore reflects nothing about the causation of the system (assuming a set of $\sigma$ values ordered by descending values). The average of the remaining values, here called $\gamma{*}$, closer approximates the causal primitives of a system. This can be seen via the same simulations previously detailed in S1 of manipulating a system along increasing axes of uncertainty in its effects (via noise), as well as uncertainty in its causes (via common causes). For the same manipulation in the same system as S1, Fig. S3A plots the original state-averaged dynamical reversibility $\gamma$ from \cite{zhang2025dynamical} against the new $\gamma{*}$, and shows how they behave compared to the causal primitives. $\gamma{*}$ behaves very similarly to the determinism plus specificity (including, e.g., becoming zero when the system is at full randomness, while $\gamma$ does not).

These adaptations put the SVD method more in line with the CE 2.0 framework. In this adaptation of the SVD method for CE 2.0, the total amount of causal emergence can be estimated as the highest-available gain that is non-trivial: $\sigma_2 -\gamma{*}$. When calculated in the “block model” system as in Figure. 4 of the main text, across the same scheme of probability redistribution, this value behaves similarly to the total gain in CP (plotted in Fig. S3B). In fact, in the initial configuration of the system prior to probability redistribution, the total gain of CP (in terms of sufficiency plus necessity) at the macroscale is actually identical to the $\sigma_2 -\gamma{*}$ value.

For comparison, the values applying the "vague" causal emergence from \cite{zhang2025dynamical} are shown. Since that method approximates the gains in EI, it inherits some of the same limitations of the CE 1.0 framework. For instance, during probability redistribution of this type within a macroscale, the measure is mathematically unstable, reducing to zero following any probability redistribution of this kind within a macrostate (see Figure S3B for a plotted value, where using a low $\epsilon$ to include most non-zero $\sigma$ values is labeled “CE 1.0 (SVD)”).

\begin{figure}[H]

    \centering
    \includegraphics[width=1\textwidth]{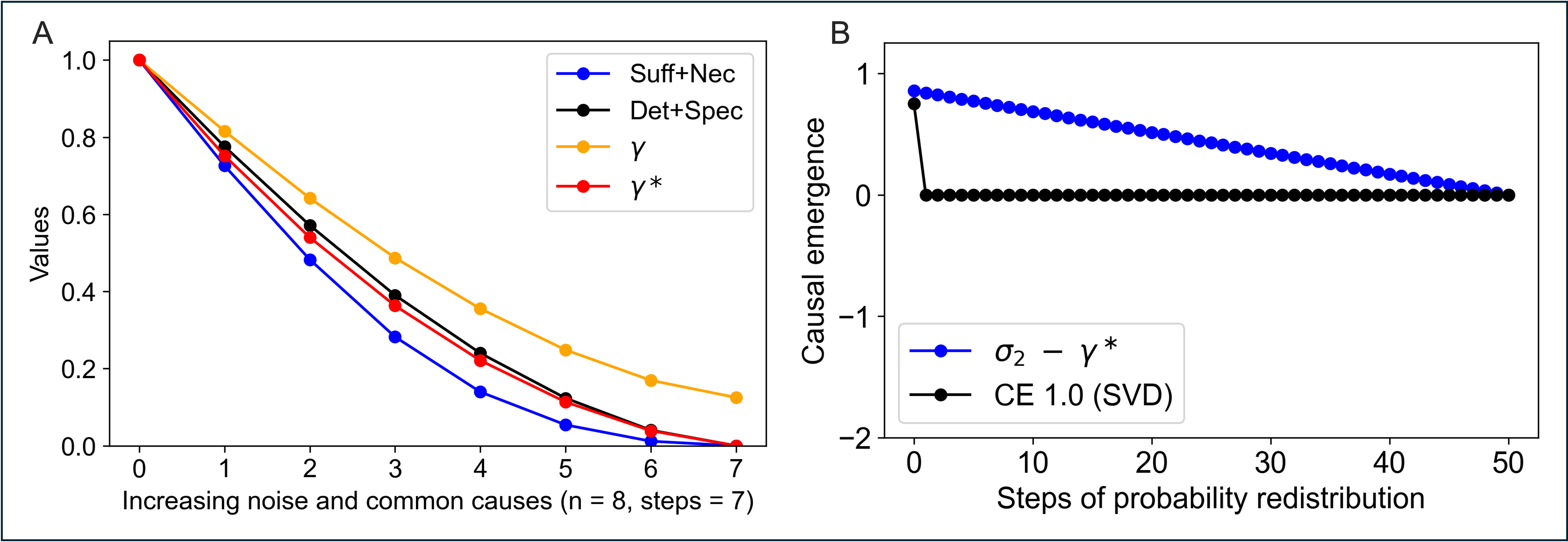} 
    \caption{\textbf{Adapting the SVD method to CE 2.0 is demonstrated across two different models of probability redistribution} (A) For the same steps of increasing noise and common causes in the same system as in S1, the CP values are again plotted, but also the proposed changes to the averaged dynamical reversibility, wherein $\gamma{*}$ is used instead of $\gamma{*}$. (B) Plotting the measures for a different case probability redistribution, this time the same as in Figure 4 of the main text. There, a "block model" consisting of two equivalency classes has its probabilities redistributed until it is entirely composed of self-loops. Plotted in black are the values when using the dynamical reversibility method detailed in \cite{zhang2025dynamical} to calculate causal emergence during this probability redistribution. Just as when using the EI, the measure detects no macroscale causation in most configurations. However, when using the adaptation of the SVD method for CE 2.0 proposed herein, the total gain ($\sigma_2 -\gamma{*}$) operates similarly to the total gain in the causal primitives (compare to Fig.4 in the main text).}
    \label{fig:2}

\end{figure}

Furthermore, applying a version of the causal apportioning schema enables an SVD-based assessment of multiscale causal structure. Specifically, each $\sigma$ (excluding $\sigma_1$) can be compared to the average of the remaining values, $\gamma*$. Causal contributions can be assessed for all $\sigma_i$ (where $2 \leq i \leq n$) that are positive, and thus satisfy:

\[
\sigma_i > \frac{1}{n-1}\sum_{j=2}^{n}\sigma_j
\]

This way of using the SVD to identify the unique causal contributions of different scales was tested on the multiscale structure of the model systems used in Section 4 of the main text. For the first system in Figure 3 of the main text (Fig.3A-C), which lacks multiscale structure, this method results in just a single positive value from the multiscale SVD analysis: 0.61, indicating a single top-heavy macroscale (agreeing with the path-based analysis from the main text). Meanwhile, the second system (Fig.3D-F), which does have mesoscale structure, has three positive values: 0.489, 0.06, 0.06, indicating the presence of one or more mesoscales.

Ultimately, these results indicates that adapting the SVD method designed for the CE 1.0 framework in \cite{zhang2025dynamical} is a promising heuristic for the CE 2.0 analysis. It has the advantage of avoiding combinatorial explosions, in that both the total gain in causal primitives at a macroscale, and the causal contributions of different scales themselves, can be estimated directly based on just the TPM of the microscale. However, it is important to note that this is just one proposal for how to adapt CE 2.0 for the SVD framework, and future research may either fully explicate this method or propose alternatives.

\bibliographystyle{unsrt}  %
\bibliography{references}  %

\end{document}